\documentclass[runningheads,a4paper]{llncs}

\usepackage{amssymb}
\usepackage{graphicx}
\usepackage{color}
\usepackage{braket}
\usepackage{cite}
\usepackage{comment}
\usepackage{gensymb}
\usepackage{booktabs}

\usepackage{url}
\urldef{\mailsa}\path|imaksymov@csu.edu.au, ivan.maksymov@gmail.com|    
\newcommand{\keywords}[1]{\par\addvspace\baselineskip
\noindent\keywordname\enspace\ignorespaces#1}

\begin{document}

\mainmatter  

\title{Cognition in Superposition:\\Quantum Models in AI, Finance, Defence, Gaming and Collective Behaviour}

\titlerunning{Cognition in Superposition}

%
%
\author{Ivan S. Maksymov}
\authorrunning{Ivan S. Maksymov}

\institute{Artificial Intelligence and Cyber Futures Institute,\\ Charles Sturt University, Bathurst, NSW 2795, Australia\\
\mailsa}

%
%

\maketitle

\begin{abstract}
At first glance, quantum mechanics and behavioural science seem worlds apart---one rooted in equations and particles, the other in thoughts and choices. Yet, emerging research reveals a profound and unexpected bridge between them. This chapter explores that bridge through quantum models of cognition and decision-making, showing how principles from quantum mechanics can help understand, and even predict, the complexities of human perception, behaviour and societal dynamics. We introduce a computationally accessible framework grounded in quantum theory, designed to model ambiguity, bias and choice in a way that classical logic cannot. Drawing on interdisciplinary sources, we demonstrate how this approach not only enriches our understanding of individual cognition but also extends to AI, video game design, financial behaviour and collective decision-making in society. Through vivid case studies---from optical illusions in video games to decision biases in economic, defence and political contexts---we show how quantum models offer new explanatory and predictive power. Thus, this chapter invites readers from all professional backgrounds to reimagine cognition and decision-making, opening new avenues for science, technology and human understanding.

\keywords{Artificial Intelligence (AI), Backfire Effect, Defence, Financial Decision-Making, Human Behaviour, Machine Vision, Neural Networks, Neuroscience, Opinion Polarisation, Optical Illusions, Politics, Psychology, Quantum Cognition, Quantum Mechanics, Social Networks, Video Games}
\end{abstract}

\begin{quote}
    `{\it Madness is something rare in individuals---but in groups, parties, peoples, epochs it is the rule.}' \\
    \hfill \textsc{Friedrich Nietzsche}
\end{quote}

\section{Introduction}
In the late 19\textsuperscript{th} century, many scientists believed that physics had nearly exhausted its potential to contribute to human understanding. However, a critical error arising from classical physical theories challenged this belief and paved the way for the development of quantum mechanics~\cite{Kra00}. Throughout the 20\textsuperscript{th} century, quantum mechanics became a foundational tool that enabled scientists across disciplines to uncover new physical phenomena and design innovative technologies. Today, we are witnessing the emergence of commercial quantum technologies that promise to revolutionise daily life and work, offering capabilities far beyond those of conventional computing systems~\cite{Gri04, Ber23}.

Intriguingly, quantum mechanics has also found a place in the social sciences. Psychologists and behavioural scientists are now employing its principles to better understand how humans form beliefs, perceive their environment and make decisions~\cite{Bus12, mindell2012quantum, wendt2015quantum, Pot22, Geo_book, Khr_book}. This chapter introduces readers to the fascinating convergence of behavioural science, psychology, AI and physics, illustrating the advantages quantum models offer over traditional approaches to understanding human perception, cognition and decision-making. We also show how the foundational principles of quantum theory can be applied to the development of games, virtual reality and machine vision systems that train individuals, such as astronauts, pilots and autonomous vehicle operators~\cite{Yam06, Cle13, Cle17}, in topics like visual perception, object recognition, situational awareness and autonomous decision-making~\cite{Atm04, Khr06, Atm10, Bus12, Aer14, Khr18, Aer22, Kov22, Pot22}.

This chapter differs in structure from most existing works, which tend to adopt a singular perspective---be it psychological, mathematical, computational or data-driven. Instead, we focus on the applied concepts of quantum mechanics and their relevance to key behavioural patterns. While some familiarity with physics and mathematics is helpful, we aim to explain concepts in the accessible language typically used in high school science education. We also provide access to a simple, modifiable computational code, enabling readers to experiment with and develop their own quantum-inspired behavioural models.
\begin{figure}[t]
\centering
\includegraphics[width=1.25\columnwidth]{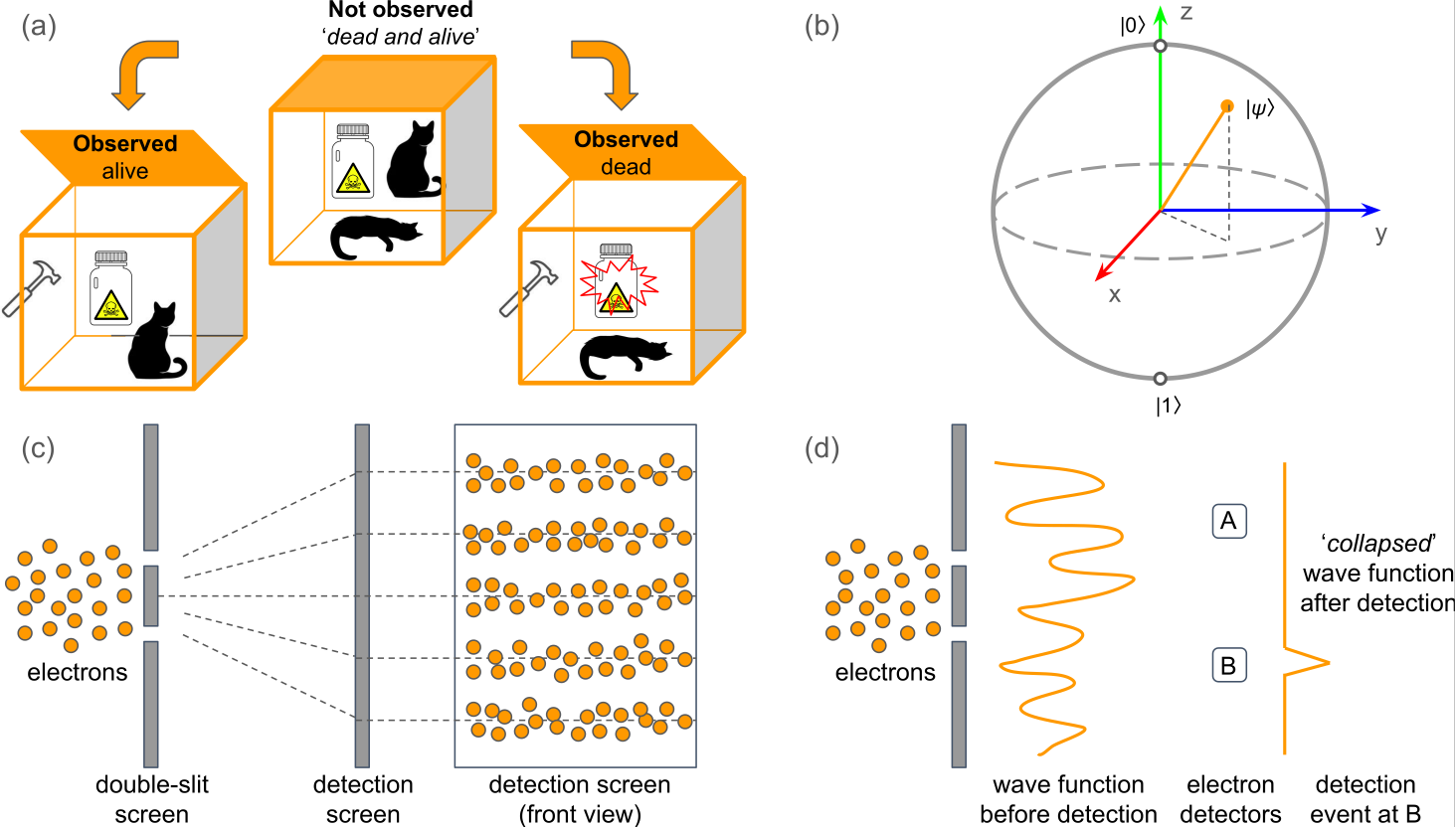}
\caption{{\bf(a)}~Schr{\"o}dinger's cat thought experiment. A cat is placed in a sealed, opaque box containing a radioactive atom, a Geiger counter, a vial of poison and a hammer. If the atom decays, the Geiger counter triggers the hammer to release the poison, killing the cat. Until the box is opened and observed, quantum mechanics suggests the cat exists in a superposition of being both alive and dead. {\bf(b)}~Illustration of a projective measurement of a qubit $\ket{\psi}$ using the Bloch sphere, where measurement collapses the qubit from a superposition to a definite state. {\bf(c)}~The double-slit experiment, which demonstrates quantum interference, showing that particles such as electrons behave as waves, creating an interference pattern until observed. {\bf(d)}~Illustration of wavefunction collapse triggered by detection using an electron detector.}
\label{Fig1}
\end{figure}

\section{Quantum mechanics and human behaviour}
Mechanics, a foundational domain of classical physics, investigates the motion of physical objects, such as balls, vehicles and celestial bodies, under the influence of forces. In contrast, quantum mechanics governs the behaviour of matter and energy at atomic and subatomic scales, where classical descriptions fail to apply~\cite{Gri04}. Also referred to as quantum theory, it provides a framework for understanding phenomena involving particles like photons and electrons that defy classical intuition~\cite{Gri04}. Beyond its theoretical depth, quantum mechanics underpins a wide array of modern technologies, including semiconductor devices, medical imaging systems, optical fibre communication networks and the emerging field of quantum computing~\cite{Nie02, Ber23}.

Quantum mechanics resists intuitive understanding when approached through the lens of classical physics~\cite{Gri04}. Among its most perplexing features is the principle of superposition, wherein a quantum system can exist simultaneously in multiple possible states until observed---a concept famously illustrated by Schr{\"o}dinger's cat thought experiment (Figure~1a)~\cite{Sch35, Sch49, Gri04}. Another cornerstone of quantum theory is Heisenberg's uncertainty principle that asserts a fundamental limit to the precision with which pairs of physical properties, such as position and momentum, can be known simultaneously~\cite{Gri04}. In practice, increasing the precision of one measurement inherently decreases the accuracy with which the other can be determined. Further expanding this conceptual framework is the phenomenon of quantum entanglement, whereby two or more particles become correlated in such a way that the state of each particle cannot be described independently of the others, even when separated by vast distances.

While these principles will be discussed in more detail later, we are already in a position to observe striking analogies between quantum mechanics and certain aspects of human behaviour. Much like a quantum system in superposition, an individual's behaviour may span a range of potential states. The same person can act in markedly different ways depending on context, suggesting that behaviour as a whole cannot be fully observed at once and that only specific facets become manifest in particular situations~\cite{Pat11}. Furthermore, when a behavioural study is designed to isolate and measure one psychological trait with high precision, other traits may become obscured or less reliably measurable, mirroring the logic of the uncertainty principle~\cite{Fle10}. This analogy offers a compelling reinterpretation of human cognition and personality through the probabilistic nature of quantum theory~\cite{Khr06, Bus12, Pot22, Khr_book}.

The influence of the environment on behaviour has long been a subject of philosophical inquiry, notably captured in Ortega y Gasset's famous assertion:~`I am myself and my circumstance'~\cite{Ortega_Gasset}. This concept of circumstance has more recently been connected to quantum aspects of cognition~\cite{deC13, Mak24_information, Mak24_information1}. Representing human behaviour as a quantum system aligns with the notion of discrete mental states~\cite{Aer22, Aer22_1, Mak24_information, Mak24_information1} and with the view of information as corresponding to energy states within physical systems~\cite{Dit14}. As we will demonstrate, these physical principles can offer insights into variations in behaviour among individuals and social groups, potentially enabling new approaches to enduring philosophical questions concerning morality, truth and values in cultural and societal contexts~\cite{Nie86}.

Earlier research showed that situation-specific behavioural patterns can be analysed using models based on classical physics~\cite{Gal04, Gal_book}. While these classical approaches have been practically effective, attempts to refine them have led to the hypothesis that classical mechanics alone is insufficient to explain consciousness and the complexity of the human cognition. Instead, quantum-mechanical processes have been proposed as a more suitable model~\cite{Khr06, Khr18, Bus12, mindell2012quantum, wendt2015quantum, Pot22, Geo_book, Khr_book}. Although the theory of the quantum cognition remains largely philosophical and mathematical in nature, empirical applications of quantum mechanics to areas such as optical illusions~\cite{Kor05, Atm10, Bus12, Ben18, Mak24_illusions, Mak24_APL}, paradoxes in decision-making~\cite{Yuk10, Aer12, Mar13, Mar16, Wei19, Sur19, Mor20} and models of social and political behaviour~\cite{Tes15, All18} have shown promise. These studies demonstrate that quantum models not only replicate results previously achieved with classical approaches but also introduce additional degrees of freedom, allowing for more accurate and elegant descriptions of human behaviour.

Demonstrations of the agreement between the predictions of quantum mechanical and classical models of human behaviour align with the core principles of a prominent theory in quantum physics known as Quantum Darwinism. This theory seeks to explain the emergence of the classical world from the quantum realm through a process analogous to Darwinian natural selection~\cite{Zur09, Zur15}. In quantum mechanics, decoherence refers to the alteration of a quantum system's state due to its interaction with the environment. As a result, the system becomes entangled with its surroundings in a way that prevents it from being described independently. According to Quantum Darwinism, it is this interaction with the environment, and not the act of observation, that causes decoherence. For example, environmental influences, rather than measurements, account for the absence of observable quantum states in large objects.

While a detailed evaluation of the advantages and limitations of Quantum Darwinism~\cite{Zur15} is beyond the scope of this discussion, the theory offers a compelling framework for understanding why, under specific conditions, we observe only certain aspects of behaviour~\cite{Sur21_1, Sur24}. In such cases, classical physics may provide sufficiently accurate descriptions. However, if the aim is to understand behaviour in its entirety, a quantum-mechanical approach becomes necessary. This is because overall behaviour arises from a superposition of quantum-like mental states and, as such, can only be adequately described using the tools of quantum mechanics.

To further contextualise this discussion, we compare the operation of a traditional digital computer with that of a quantum computer \cite{Nie02, Mak24_illusions}. A classical digital computer relies on bits, which are always in one of two discrete physical states, representing the binary values `0' and `1'. This behaviour is analogous to an on/off light switch. In contrast, a quantum computer employs quantum bits (qubits), which can occupy the states $\ket{0}$ and $\ket{1}$, analogous to the binary states of a classical bit. However, a qubit can also exist in a superposition of these states, represented mathematically as $\ket{\psi} = \alpha\ket{0} + \beta\ket{1}$, where the coefficients $\alpha$ and $\beta$ are complex numbers that satisfy the normalisation condition $|\alpha|^2 + |\beta|^2 = 1$.

From a physical perspective, the state of a qubit can be visualised on the Bloch sphere (Figure~\ref{Fig1}b). When a closed qubit system interacts in a controlled manner with its environment, measurement reveals the probabilities of finding the qubit in either of its basis states. Specifically, for the state $\ket{\psi} = \alpha\ket{0} + \beta\ket{1}$, the measurement probabilities are given by $P_{\ket{0}} = |\alpha|^2$ and $P_{\ket{1}} = |\beta|^2$. This implies that, upon measurement, the qubit collapses to one of its basis states $\ket{0}$ or $\ket{1}$. Visually, this measurement process corresponds to projecting the qubit state onto one of the coordinate axes of the Bloch sphere (e.g., the $z$-axis in Figure~\ref{Fig1}b).
\begin{figure}[t]
\centering
\includegraphics[width=1.25\columnwidth]{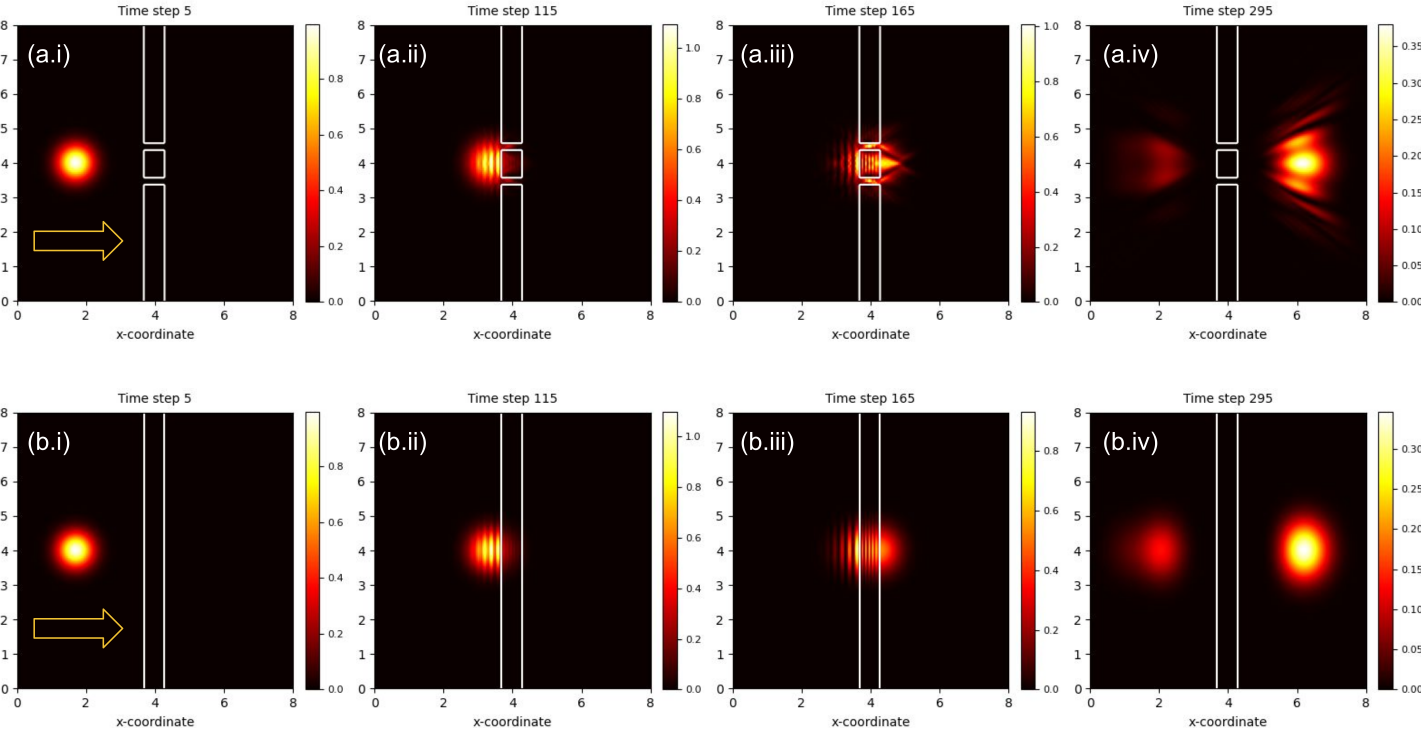}
\caption{{{\bf(a.i)--(a.iv)}~Instantaneous snapshots of an energy wave packet modelling the interaction of an electron with a double-slit structure. \bf(b.i)--(b.iv)}~Instantaneous snapshots of an energy wave packet modelling the tunnelling of an electron through a continuous potential barrier. The false-colour scale of the images encodes the computed probability density values.}
\label{Fig2}
\end{figure}

\section{Quantum physics-driven model of human behaviour}
The interference fringes observed in the famous physical double-slit experiment provide a well-known illustration of the superposition principle~\cite{Wal21}. (Other well-known physical systems can be employed without fundamentally altering the main direction of this discussion~\cite{Ben18, Sur24}.) After outlining the fundamental physics underlying this experiment, we will adopt the double-slit configuration as a conceptual model for understanding human behaviour.

\subsection{Double-slit experiment}
Atoms are the basic building blocks of matter. However, towards the end of the 19\textsuperscript{th} century, the concept of atoms was rather controversial~\cite{Smi20}. In an attempt to test the existence of atoms, Max Planck studied the properties and behaviour of blackbody radiation (a blackbody is an object that absorbs all radiation falling on it). Planck concluded that a blackbody absorbs and re-emits radiation in hypothetical discrete bits that he called \textit{quanta}~\cite{Kra00}.

In 1905, Albert Einstein provided further arguments in favour of the existence of the quanta, suggesting that radiation itself comes in discrete bits of energy called \textit{photons}. Einstein's hypothesis posed a problem since there was a well-established body of experimental evidence supporting a wave theory of light. Then, in 1923, Louis de Broglie demonstrated a direct mathematical relationship between an electron's wave-like property (wavelength) and a particle-like property (momentum). Later, in 1926, Erwin Schr{\"o}dinger used the classical theory of waves and some quantum conditions derived from de Broglie's relationship. The result was the Schr{\"o}dinger wave equation, where the motion of a particle such as an electron is calculated from its wave function~\cite{Gri04}.

The \textit{wave function} is a mathematical description of the quantum state of an isolated quantum system. In classical mechanics, we can readily interpret the theory's concepts (e.g., energy, momentum) and relate them to the physical objects that possess these quantities~\cite{Gri04}. However, in quantum mechanics, the interpretation of these concepts requires performing specific mathematical operations on the quantum state of the particle. These operations can be regarded as mathematical algorithms that yield observable values from the wave function.

To demonstrate that light behaves as a wave, scientists illuminate a narrow slit made in an opaque screen. The light passing through the slit bends around the edges and spreads out, a process called \textit{diffraction}. When the screen has two narrow slits, the light waves diffracted by the slits interfere, producing an alternating pattern of light and dark bands known as \textit{interference fringes}. This experiment can also be replicated using waves of a different nature, such as water waves.

The same experimental setup can be used with electrons (see Figure~\ref{Fig1}c). Each electron passing through the slits is registered on the screen as a single bright spot. As more and more electrons pass through, these spots begin to group together and form a double-slit interference pattern of alternating bright and dark fringes---similar to what is observed with optical waves. Each individual electron thus behaves like a wave, described by a wave function $\psi$, passing through both slits simultaneously and interfering with itself before striking the screen.

The square magnitude of the wave function, $|\psi|^2$, represents the \textit{probability density} of the particle. The alternating peaks and troughs of the wave function translate into a probability pattern: bright fringes correspond to high probabilities of finding the next electron and dark fringes correspond to low probabilities. Before the electron strikes the screen, it has a probability of being found \textit{anywhere} that $|\psi|^2 > 0$. This ability of many states to exist simultaneously exemplifies \textit{quantum superposition}.

We now return to the analogy between quantum mechanics and human behaviour. Observation of a particular aspect of human behaviour under certain circumstances, or analysis of human behaviour as a whole, can be modelled similarly to the double-slit experiment. The state of the electron is not defined until the wave function interacts with the screen, at which point it \textit{collapses}, and the electron appears at a specific location. We can say the same about human behaviour:~it is not defined until it is measured under certain conditions.

Thus, in our analogy, the double-slit structure represents the environment with which the individual interacts and the detector screen represents the particular circumstance or context that shapes observable behaviour. Several specific examples of such modelling approaches are presented below.
\begin{figure}[t]
\centering
\includegraphics[width=0.5\columnwidth]{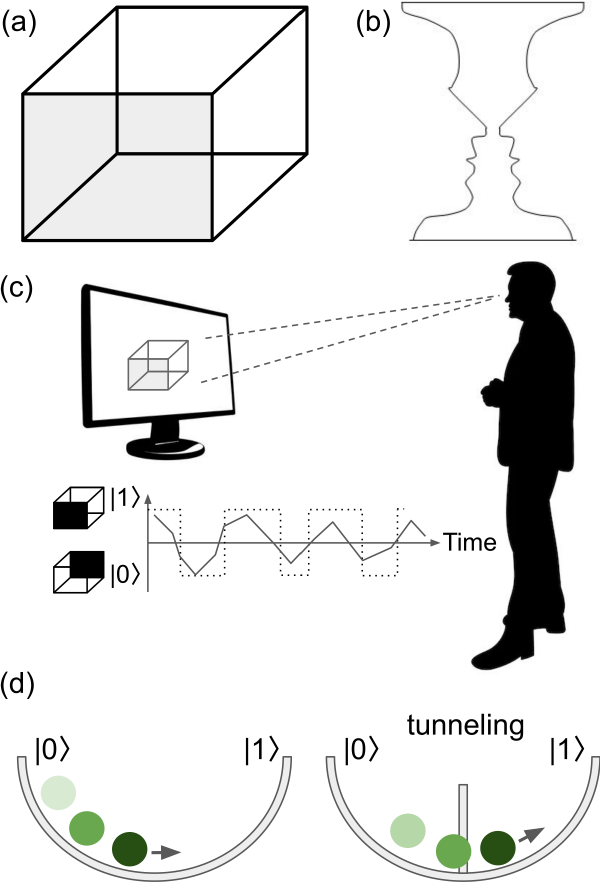}
\caption{\textbf{(a)}~\textit{The Necker cube}: When asked whether the shaded face of the cube is at the front or rear, observers will spontaneously alternate between two stable perceptual interpretations:~front ($\ket 0$) and rear ($\ket 1$). \textbf{(b)}~\textit{Rubin's vase}: Do you see two people facing each other ($\ket 0$) or a vase ($\ket 1$)? Perception flips between these two interpretations. \textbf{(c)}~According to traditional theories, perceptual switching is binary (dotted line), transitioning directly between $\ket 0$ and $\ket 1$. Quantum cognition theory suggests that observers may temporarily perceive a quantum-like superposition of both states (solid curve). \textbf{(c)}~An electron in a parabolic well behaves as a quantum harmonic oscillator and can tunnel through a barrier. The states $\ket 0$ and $\ket 1$ represent the Necker cube's fundamental percepts in the quantum oscillator model discussed in this work.}
\label{Fig3}
\end{figure}

\subsection{The model}
We numerically solve the Schr{\"o}dinger equation in two-dimensional ($x,y$-coordinate) space. This equation is a linear partial differential equation that governs the wave function of a quantum-mechanical system. Its understanding requires familiarity with the concepts and notations of calculus, including derivatives with respect to space and time. However, in the following, we will avoid discussing such mathematical concepts. Instead, we employ the Crank--Nicolson method, a well-known computational algorithm for solving certain classes of partial differential equations~\cite{Tho95}. A Python code that implements this algorithm can be accessed following the instructions given at the end of this document. Additional technical details can be found in the literature~\cite{Tho95}. Similar numerical techniques such as the finite-difference time-domain method may also be used~\cite{Sullivan}.

In brief, we use a two-dimensional spatial grid that consists of $N$ points along the $x$-direction and $N$ points along the $y$-direction. We also consider $N_t$ time points. The continuous wave function $\psi$ describing the system at a given spatial point and time is approximated as a discrete function $\psi^k_{i,j}$, where $i$ and $j$ are the spatial indices, and $k$ is the time index. Using the two-dimensional mesh and discretisation in time (i.e.,~time is represented as a sequence of~$N_t$ discrete instances), we approximate the spatial and temporal derivatives in the Schr{\"o}dinger equation using Euler's method~\cite{Tho95}. After algebraic manipulations, we obtain a matrix equation that can be solved using standard computational procedures available in scientific and technical programming languages such as Python.

Python is a programming language that, through specialised and widely used libraries such as NumPy and Matplotlib, enables matrix manipulation for implementing algorithms, processing data and plotting functions. In our algorithm, we represent an electron as an energy wave packet with a two-dimensional Gaussian (bell-shaped) profile. The double-slit structure is represented as a potential barrier for the electron. The height of the barrier is a tunable parameter that allows for custom simulation scenarios. The barrier can be removed from the model by setting its height to zero. The boundaries of the two-dimensional space act as infinitely high potential barriers, meaning the electron cannot escape the computational domain.

As a first test, we use the model to simulate a wave packet moving from the left side of the domain toward the double-slit structure (Figure~\ref{Fig2}a). Upon interaction with the slits, the packet forms an interference pattern near the right boundary. Readers can rerun the simulation using different values for the potential barrier height and observe how the interference pattern changes (in some cases, the wave packet is fully reflected). A similar effect occurs for potential barriers of different widths. One can also remove one slit or add more to generate more complex interference patterns.

As a second test, we remove the double-slit structure to form a solid barrier (Figure~\ref{Fig2}b). The simulation shows that the initial wave packet strikes the barrier and then splits into two components:~one reflected from the barrier and the other transmitted through it. This scenario illustrates the quantum-mechanical phenomenon of electron tunnelling through a potential barrier. The significance of quantum tunnelling in our main discussion will become clear in the following sections.

How can these physical results be interpreted within the framework of quantum cognition theory? The wave packet represents a single particle modelling an individual's behaviour. The associated wave function may be interpreted as a mental wave function and the observed interference can be understood as interference of the mind~\cite{Khr06}. Our model thus supports the idea that an individual's mind can interfere with itself, as proposed in earlier quantum mind hypotheses. This result is consistent with mainstream theories in psychology~\cite{Bom20, Qur20} and philosophy~\cite{Dic93}.

It is straightforward to modify the computational code to introduce a second individual, modelled by a different wave packet. For example, a second energy wave packet may move from right to left, colliding with the first in the middle of the domain. This setup produces interference between the two wave packets. Similar simulations involving three or more wave packets can demonstrate multi-mind interference. One can also introduce a potential structure (e.g., a double-slit) to study the effect of environmental influences on individual and collective behaviour. These simulations provide a basis for investigating both interpersonal interactions and environmental effects.

Interestingly, the core physical ideas illustrated by the simple yet powerful model in this section resemble those used in well-known quantum computing approaches. One example is the well-known Shor's algorithm \cite{Sho94} (quantum algorithms like Shor's can solve factorisation problems in seconds, a task that might take even the fastest classical computers millions of years \cite{Nie02}). Another is the quantum neural network architecture proposed by Menneer and Narayanan \cite{Men95, Nar00}. 

In the Menneer--Narayanan framework, a memory register is first placed into a quantum superposition, meaning it simultaneously holds all possible integer values. Each of these values then evolves independently in what can be thought of as separate `branches' or `paths', a concept that some researchers relate to the many-worlds interpretation of quantum mechanics \cite{Ezh00, Eve57, Eve57_1, Whe57, sep-qm-manyworlds}. Computation proceeds along each of these paths in parallel and the process ends when these quantum branches begin to interfere with one another. The resulting interference patterns, which resemble standing waves, help reveal repeating patterns linked to the factors of the original number.

Although this method does not always yield the correct answer on the first try, it can be efficiently repeated and the result verified. Beyond number factoring, similar quantum principles have found intriguing applications in quantum cognition theory, where the superposition of states is used to represent the multitude of opinions or beliefs held by an individual or across a social network. Just as quantum algorithms explore many possible solutions simultaneously, cognitive models inspired by quantum theory propose that a person's mind may hold overlapping and sometimes contradictory perspectives, which only `collapse' into a specific decision or belief when a choice or judgement is required.
\begin{figure}[t]
\centering
\includegraphics[width=0.7\columnwidth]{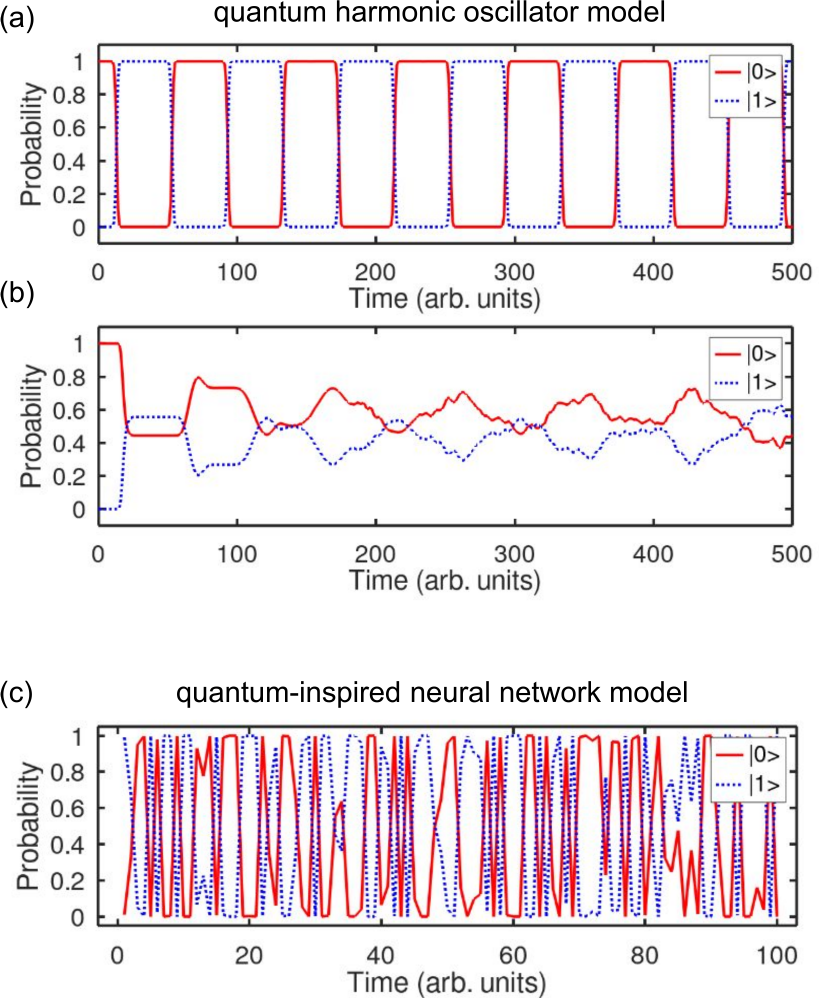}
\caption{Results from the quantum oscillator model:~\textbf{(a)}~without the potential barrier and \textbf{(b)}~with the potential barrier included. Data points with probabilities $P_{\ket 0} = 0$ or $P_{\ket 1} = 1$ represent the fundamental perceptual states of the Necker cube. Intermediate points, where $0 < P_{\ket 0}, P_{\ket 1} < 1$ and $P_{\ket 0} + P_{\ket 1} = 1$, correspond to superpositions of the perceptual states $\ket 0$ and $\ket 1$. \textbf{(c)}~Prediction of the Necker cube perception made by the quantum neural network. The time units used in panels (a) and (b) are different.}
\label{Fig4}
\end{figure}

\section{Illustrative applications of the quantum model}
\subsection{Optical illusions}
Optical illusions have long captivated both scientists and the general public~\cite{Nek32, Was34, Sha17}. They have also served as inspiration for artists, photographers and advertisers~\cite{Fis67, Lin01}. Ambiguous images such as the Necker cube~\cite{Nek32} and Rubin's vase~\cite{Par08, Kha21} (Figure~\ref{Fig3}a,~b) as well as \emph{My Wife and My Mother-in-Law}~\cite{Nic18}, the Rabbit–Duck illusion~\cite{McM10} and the Spinning Dancer illusion~\cite{Tro10} have been paradigmatic tools in psychological and psychiatric research~\cite{Lon04, Kor05, Bus12, Sto12, Mei19, Bas16}.

Beyond psychology, optical illusions are of growing interest in the development of computer vision and robotic systems that seek to interpret the world in a human-like manner~\cite{Tak13, Mat18}. Realistic models of optical perception are also used to enhance video game experiences and virtual reality technologies~\cite{wang2021game, Li15, Neckerworld, game1, game2, Mak24_illusions, Mak24_APL} as well as they play an important role in pilot and astronaut training~\cite{Yam06, Cle13, Cle17}.

Recently, researchers exploring quantum models of cognition, perception and decision-making have turned their attention to optical illusions~\cite{Atm10, Bus12, Pot22}. The Necker cube (Figure~\ref{Fig3}a), in particular, has been a canonical example for understanding bistable perception~\cite{Kor05}. It demonstrates how the human brain can switch between two equally plausible interpretations of the same stimulus, modelled as two stable states~\cite{Ino94, War15}.

In a typical Necker cube experiment, an observer is asked:~`Is the shaded face at the front or at the rear?' Using a binary input device, such as an electronic pushbutton, responses are recorded over time and plotted as a sequence of square pulses that alternate between two states, commonly labelled $\ket 0$ and $\ket 1$ (Figure~\ref{Fig3}c, dotted line). The duration of these states varies between individuals and depends on factors such as age and gender~\cite{lo2011investigation}, yet the general bistable pattern remains consistent across studies~\cite{War15, Wan17_1}.

Simultaneous recordings of brain activity and eye movements during such experiments have revealed a deeper complexity~\cite{Pia17, Joo20}. Analysis suggests that perception may not simply alternate between $\ket 0$ and $\ket 1$, but instead reside in a superposition of both states (Figure~\ref{Fig3}c, solid line). That is, even when a conscious decision is made to report a single state, neural and ocular data imply concurrent awareness of both interpretations, an idea reminiscent of the famous Schr{\"o}dinger’s cat thought experiment.

To describe this phenomenon, Busemeyer and Bruza~\cite{Bus12} proposed a quantum model of bistable perception based on the Schr{\"o}dinger equation. Unlike classical Markov models which constrain transitions strictly between $\ket 0$ and $\ket 1$, their quantum approach allows the system to oscillate between the two, enabling superposition. A similar conclusion was independently reached by Atmanspacher and Filk~\cite{Atm10} via their quantum `Necker-Zeno' model.

Harmonic oscillators are fundamental in physics and appear in mechanical, electrical and electromagnetic systems. A classical harmonic oscillator, like a marble rolling in a bowl (Figure~\ref{Fig3}d, left), can be described using a parabolic potential. However, when quantum effects become significant, one must solve the Schr{\"o}dinger equation for a particle in such a potential. The solutions yield discrete energy levels
\begin{equation}
E_n = \left(n + \frac{1}{2}\right)\hbar\omega,\quad n = 0, 1, 2, \ldots \,,
\end{equation}
where $\hbar$ is the reduced Planck constant and $\omega$ is the angular frequency.

Building on this, Maksymov and Pogrebna~\cite{Mak24_information, Mak24_illusions} extended the quantum model of Necker cube perception by incorporating quantum tunnelling. While a classical particle (e.g., a marble) cannot cross a high barrier placed in the bowl, a quantum particle such as an electron can tunnel through it (Figure~\ref{Fig3}d, right). This tunnelling behaviour captures transitions in perception that defy simple binary switching and aligns more closely with data from brain activity and eye tracking~\cite{Ein04, Joo20, Ang20, Cho20, Mat23}.

Figure~\ref{Fig4}a,~b shows the results of simulations conducted using a parabolic potential well and a parabolic well with a barrier, respectively. In both cases, the potential well was virtually split into two equal spatial regions, denoted as~$\ket 0$ and~$\ket 1$. The probabilities of finding the electron in each of these regions were calculated and interpreted as the probabilities of perceiving the Necker cube in the corresponding perceptual states.

The model with the parabolic well effectively reproduces the results obtained using a simpler quantum model proposed by Busemeyer and Bruza~\cite{Bus12}, where perception alternates between the~$\ket 0$ and~$\ket 1$ states, interspersed with intervals when perception is in a superposition of these states. This behaviour is also consistent with the findings of Atmanspacher and Filk~\cite{Atm10}. Moreover, introducing a barrier into the potential well and incorporating quantum tunnelling effects allowed for a more accurate model, with improved agreement with predictions based on experimental brain activity and eye-tracking data.

The results shown in Figure~\ref{Fig4}b are also consistent with those reported in previous work~\cite{Wil23}, which demonstrated that perception can become unstable before a perceptual reversal is consciously reported. Experimental evidence further shows that observers can unconsciously influence the number of perceptual reversals over time~\cite{Lon04}, although complete prevention of reversal is not possible, a phenomenon also captured by the model.

To validate the quantum oscillator model, a deep neural network digital twin was implemented~\cite{Mak24_illusions}. The network architecture consisted of an input layer with 100 nodes, three hidden layers with 20 nodes each and an output layer with two nodes representing the~$\ket 0$ and~$\ket 1$ states. The input layer encoded an image of the Necker cube, while the output nodes were used to classify the perceived state.

Although the physical basis of perceptual switching remains under investigation~\cite{Kor05}, one prominent theory attributes it to chaotic dynamics found in nonlinear physical and mathematical systems~\cite{Ino94, Shi10, Che23}. The fact that the brain functions as a complex, nonlinear, and chaotic dynamical system~\cite{Mck94, Kor03} supports this hypothesis.
\begin{figure}[t]
\centering
 \includegraphics[width=0.7\textwidth]{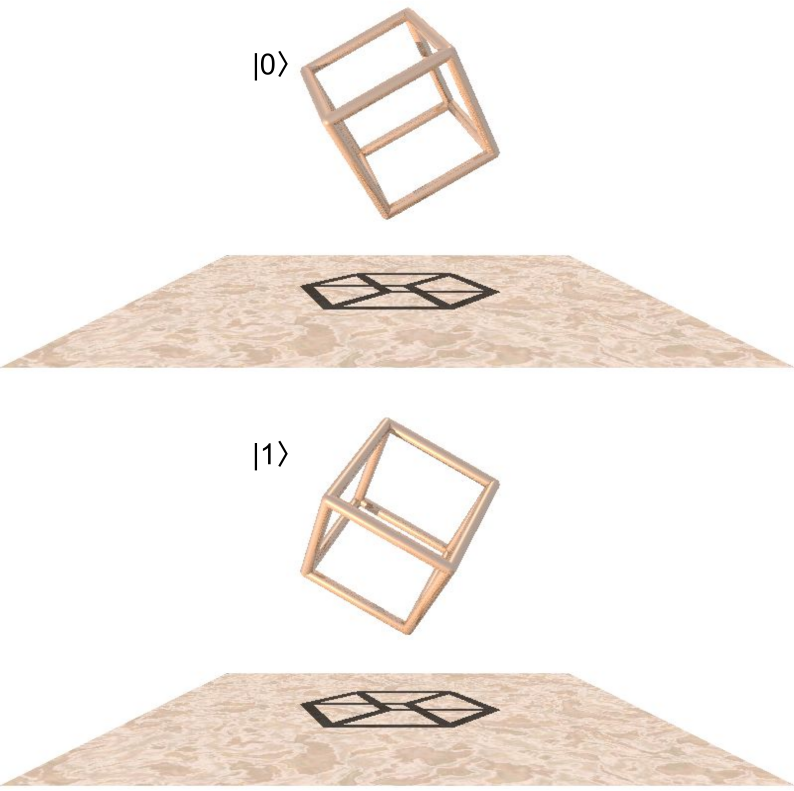}
 \caption{Projective qubit-like measurement applied to the Necker cube: the two-dimensional shadows, perceived as an ambiguous Necker cube, are treated as a qubit-like superposition of $\ket 0$ and $\ket 1$. By virtually projecting these shadows back into three-dimensional space, we obtain an unambiguous cube corresponding to one of the basis states.\label{Fig4_1}}
\end{figure}

To emulate chaotic dynamics in the neural model, a quantum-physical true random number generator was used to define the network's internal connections~\cite{Sym11}. Unlike pseudo-random number generators commonly used in computational software, true random numbers are produced in real time by measuring quantum fluctuations of the vacuum. While classical physics views the vacuum as empty, quantum physics describes it as filled with electromagnetic field fluctuations across all frequencies. These fluctuations are measured and converted into random numbers, which are either broadcast online or generated by supercomputers~\cite{Kum23}. Regardless of the source, neural networks configured using quantum-generated randomness are unbiased toward either perceptual state and produce non-repeating prediction sequences~\cite{Mak24_illusions}.

Figure~\ref{Fig4}c illustrates the behaviour of the quantum neural network, which produces a chaotic switching pattern between the~$\ket 0$ and~$\ket 1$ states of the Necker cube. This suggests that the perceptual state can also exist in superposition. Notably, the predictions from the neural network model are qualitatively consistent with those from the quantum oscillator model. A detailed comparison shows that the neural network pattern combines features from both oscillator variants, with and without the potential barrier. For instance, during the time interval~$T = 0$ to~$T = 30$, the neural network output resembles the quasi-periodic behaviour of the basic quantum oscillator. Between~$T = 30$ and~$T = 60$, the network's output aligns more closely with that of the oscillator with a barrier. While the timescales in the two models differ, this discrepancy is not critical and can be addressed by modifying the parabolic potential profile accordingly.

The quantum oscillator model could also be enhanced by incorporating more complex potential well geometries. For example, the simulation code accompanying this study can be extended to create two-dimensional wells with three or more valleys and/or barriers. By assigning states such as~$\ket 0$, $\ket 1$, $\ket 2$, etc., to each valley and calculating the probability of locating the electron in each one, a model of multistable perception can be developed. Such a model could be applied to study complex visual illusions, including tristable depth stimuli~\cite{Wal13}. Moreover, a multivalley well could support decision-making tasks, such as classifying facial images into social categories (e.g., `good guy' vs. `bad guy'), as discussed by Busemeyer and Bruza~\cite{Bus12}.
\begin{figure}[t]
\centering
\includegraphics[width=0.75\columnwidth]{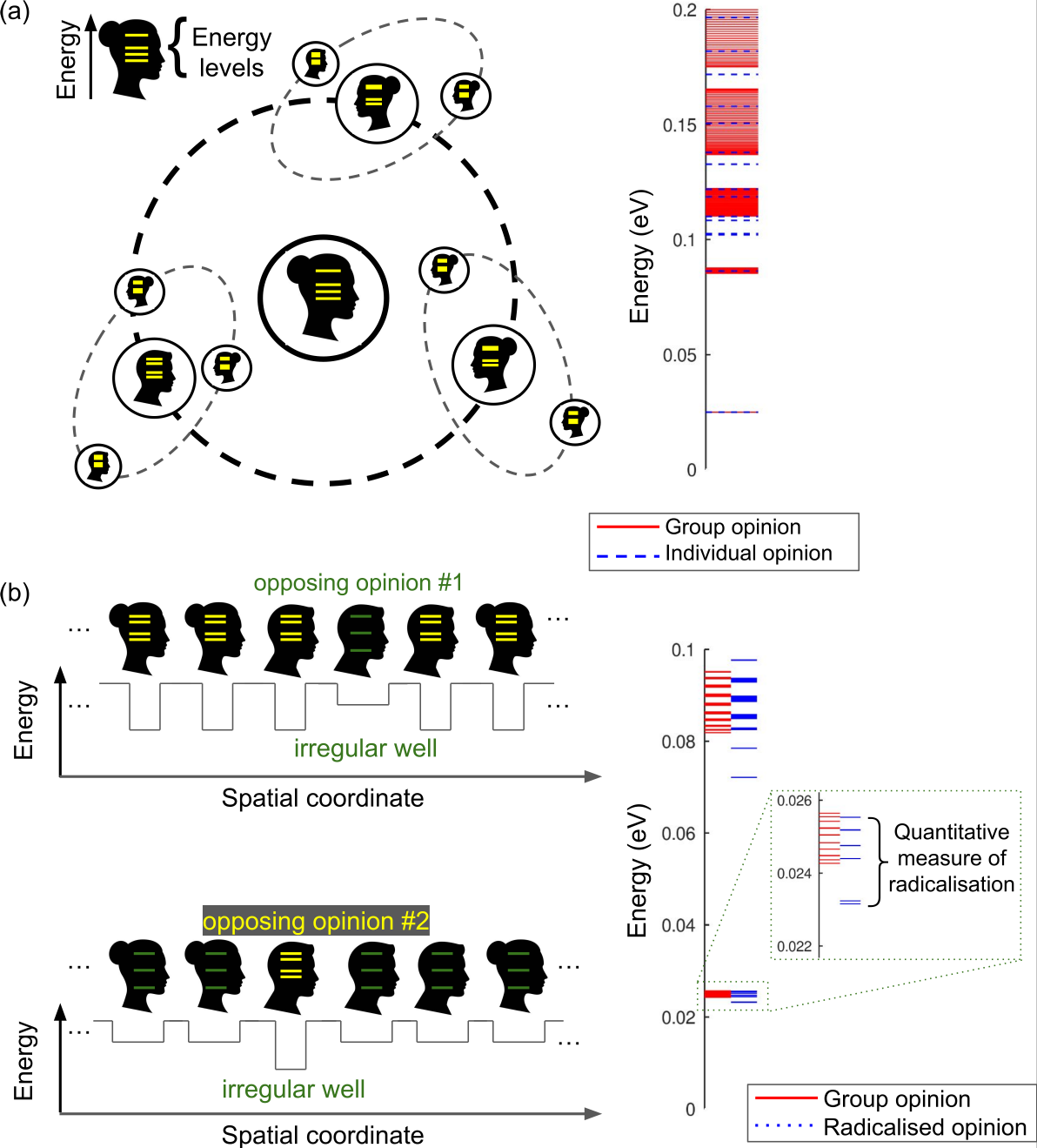}
\caption{\textbf{(a)}~A social network and individual belief systems are illustrated as discrete energy levels (horizontal lines overlaid on head silhouettes). The right panel shows simulation results: dashed lines represent an isolated individual, and solid lines depict the network. Clustering of solid lines models opinion formation. \textbf{(b)}~Opposing opinions are modelled via irregular potential wells with distinct energy level structures. Different geometries (cf.~opposing opinions \#1 and \#2) represent various opposing views. Changes in level groupings reflect opinion radicalisation.}
\label{Fig5}
\end{figure}

To conclude this section, Figure~\ref{Fig4_1}, rendered using the physical ray-tracing software POV-Ray~3.7, illustrates how the principle of projective qubit measurement (see Figure~\ref{Fig1}b and the relevant discussion in the main text) can be generalised to the Necker cube. The two-dimensional (bottom) images in Figure~\ref{Fig4_1} are shadows cast by visually distinct three-dimensional cubes. However, the shadows themselves are identical, forming an ambiguous Necker cube with alternating left and right faces (perceived after 5--10\,seconds of observation, with some observers needing to blink to notice the illusion~\cite{Ang20}). Drawing an analogy with the projective measurement in Figure~\ref{Fig1}b, we treat the shadows as a qubit-like superposition of the two fundamental perceptual states of the cube and virtually project them back into three-dimensional space to obtain an unambiguous image of the cube corresponding to either the $\ket 0$ or $\ket 1$ basis state.

\subsection{Opinion polarisation in social networks}
In the preceding discussion, we have demonstrated that a quantum-mechanical system permits only discrete, quantised energy levels, in contrast to the continuous spectrum of states available in classical systems. Building upon this foundational property and following previous works~\cite{Aer22, Aer22_1}, it has been proposed that human belief systems can be effectively modelled as discrete energy levels of a quantum system~\cite{Mak24_information}.

In the quantum harmonic oscillator model represented by a parabolic potential well, the energy levels are evenly spaced~\cite{Kittel, Gri04, Smi20}. However, in the case of a rectangular potential well, the energy states follow the relationship $E_n \propto (n/L)^2$, where $n = 1, 2, \dots$ denotes the quantum state and $L$ is the well width~\cite{Kittel, Smi20}. Therefore, by adjusting the parameter $L$, one can modulate the structure of allowed energy levels.

This physical framework can be mapped onto a model of a social network composed of interacting individuals. Let us assume that a single rectangular potential well represents one individual, and the system of beliefs of this individual is encoded in the corresponding discrete energy levels (see Figure~\ref{Fig5}a, top left inset). Adding a second individual with identical beliefs results in an identical potential well. Conversely, if the second individual holds different beliefs, we represent this using a well with altered geometry and a distinct energy spectrum.

Extending this approach, we can construct an arbitrarily large social network of individuals, each represented by a unique potential well reflecting their belief system (Figure~\ref{Fig5}a, central panel). Since this quantum social network model adheres to the laws of quantum mechanics, solving the Schr{\"o}dinger equation enables us to simulate dynamic behaviours such as opinion change, disagreement, conflict and misinformation~\cite{Ace13}.

Analogous to how atomic orbitals combine to form energy bands, when like-minded individuals interact, their belief systems overlap, resulting in grouped energy levels or `band formation'~\cite{Kittel}. This is depicted in Figure~\ref{Fig5}a (right), where dashed lines denote individual belief states, and solid lines represent emergent group beliefs~\cite{Mak24_information}. These energy levels are exact solutions of the Schr{\"o}dinger equation.

Introducing an individual with an opposing opinion (Figure~\ref{Fig5}b, left) leads to the formation of separate energy bands, representing polarised group beliefs (Figure~\ref{Fig5}b, right)~\cite{Mak24_information}. By comparing different energy band formations, we can gain insight into mechanisms of radicalisation, the spread of misinformation and the emergence of social conflict. For example, the model allows for controlled manipulation of opposing viewpoints by altering the geometry of a single potential well (so-called `irregular wells') and solving the Schr{\"o}dinger equation across the network. Readers can explore multiple irregular well shapes (e.g., parabolic, triangular or quartic) to model a broad spectrum of scenarios.

This approach has demonstrated the capability to replicate key social phenomena such as the \textit{backfire effect}~\cite{Mak24_information}, wherein individuals reinforce pre-existing beliefs upon encountering contradictory information~\cite{Swi20, Che21}. The backfire effect is particularly pronounced in beliefs tightly coupled to identity or worldview. For instance, a study revealed that Republican users exposed to Democratic content exhibited increased conservatism, while Democrats became only slightly more liberal~\cite{Bai18}. Other domains affected by this effect include vaccination~\cite{Nyh15}, climate change~\cite{Dix19} and abortion rights~\cite{Lie21}.

\begin{figure}[t]
\centering
\includegraphics[width=1.2\linewidth]{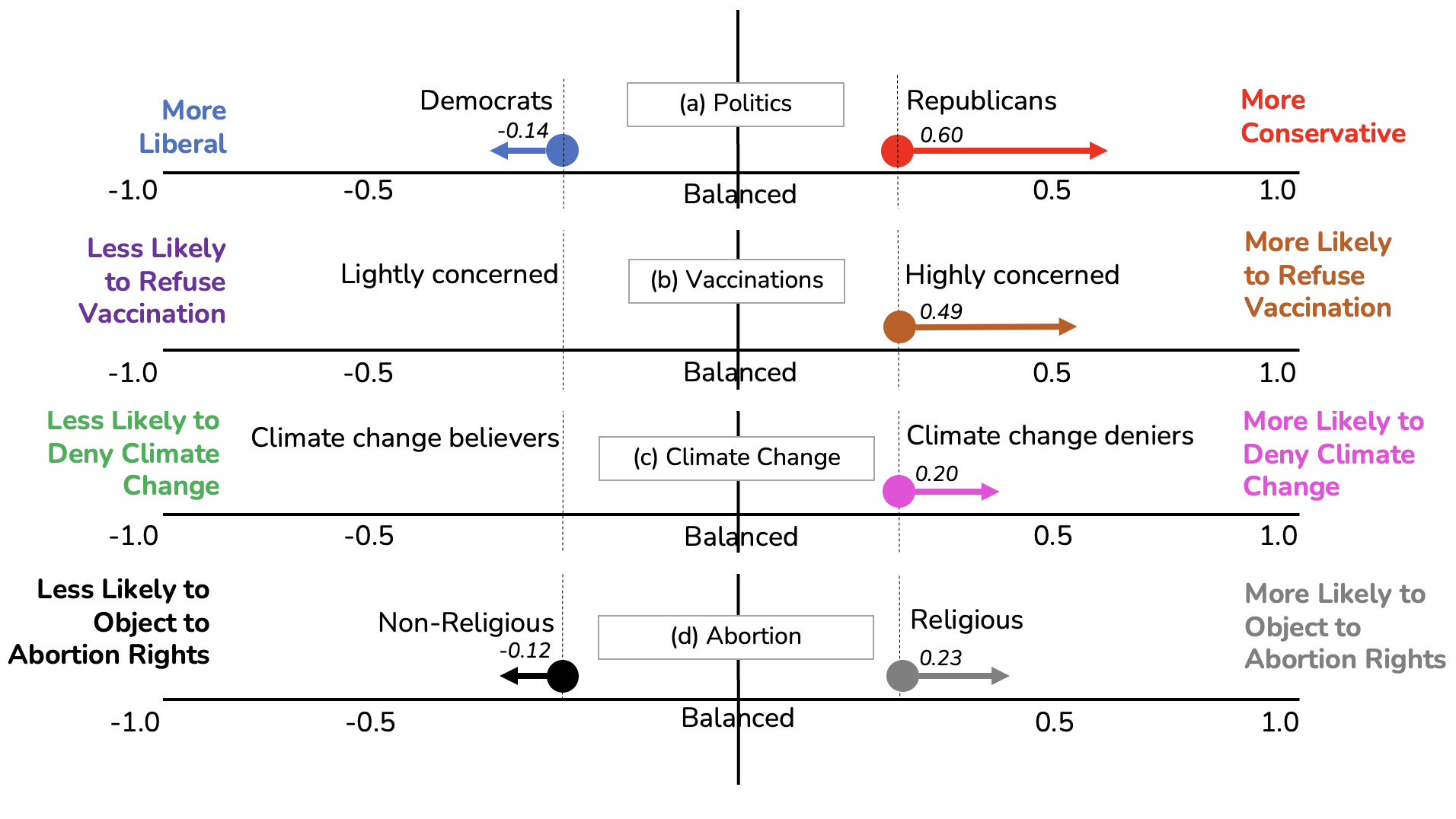}
\caption{Examples of the backfire effect reported in the literature. The figure illustrates the backfire effect observed in four studies related to: \textbf{(a)}~Politics~\cite{Bai18}, \textbf{(b)}~Vaccinations~\cite{Nyh15}, \textbf{(c)}~Climate Change~\cite{Dix19} and \textbf{(d)}~Abortion~\cite{Lie21}. In each case, exposure to opposing views strengthened pre-existing beliefs in some groups. Horizontal axes are normalised independently on a scale from $-1$ to $1$. The effect is most pronounced among Republicans, vaccine-skeptics, climate change deniers and religious individuals, who became more entrenched in their views. In contrast, Democrats and non-religious participants showed weaker or negligible backfire responses. Note: counterpart views are not available for all studies.}\label{Fig6}
\end{figure}

It is noteworthy that some domains demonstrate a `double-sided' polarisation effect (e.g., politics, abortion), while others such as climate change and vaccination exhibit a `one-sided' response (Figure~\ref{Fig6}). This discrepancy, likely due to missing data, presents challenges for computational models of opinion dynamics. However, since the quantum model can reproduce both symmetrical and asymmetrical responses, it offers a promising avenue to infer missing polarisation trends~\cite{Mak24_information}.

Furthermore, this quantum approach can extend classical sociophysical models that have been used to predict election outcomes and opinion dynamics~\cite{Gal05, Cas09, Hu17, Red19, Gal22, Int23}. For example, in the Sznajd model~\cite{Szn00} and its variations~\cite{Red19}, opinions are treated as binary spin states ($\uparrow$, $\downarrow$), and interactions only affect neighbouring agents but not the central pair. While classical methods introduce extra parameters to capture higher-order interactions~\cite{Gal97}, quantum mechanics intrinsically accounts for complex, multilateral influence across the network.
\begin{figure}[t]
\centering
 \includegraphics[width=1.2\textwidth]{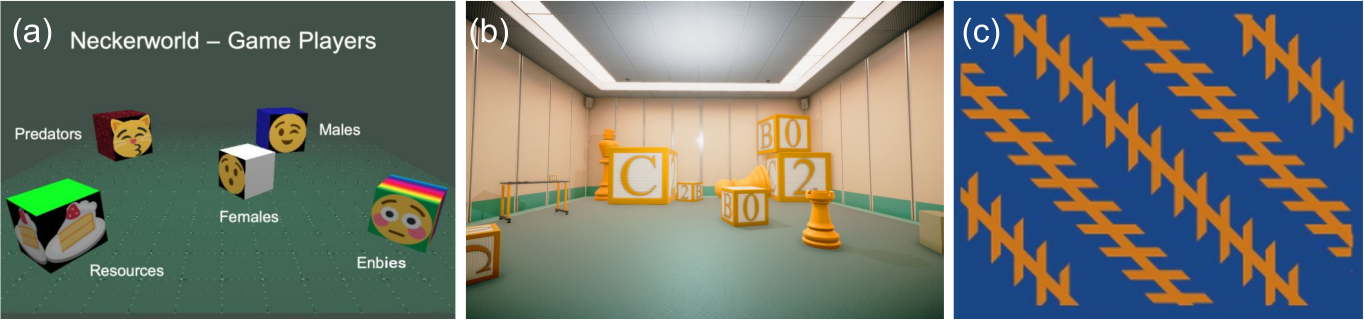}
 \caption{Screenshots exemplifying the use of optical illusions in video games: (a)~Neckerworld, an experimental computer vision game using the Necker cube illusion \cite{Neckerworld}; (b)~Superliminal, a commercial surreal puzzle game that incorporates perspective illusions \cite{game1}; (c)~an experimental video game implementing the Z{\"o}llner illusion \cite{Zol60}, illustrating the potential of visual distortions in gameplay mechanics \cite{wang2021game}.}\label{Fig7}
\end{figure}

\subsection{Decision-making in lotteries and video games}
Further practical applications of the quantum models of optical illusions, discussed in the previous sections, can be illustrated through the analysis of illusions frequently encountered in video games such as \textit{Superliminal}~\cite{game1} and \textit{Neckerworld}~\cite{Neckerworld} (Figure~\ref{Fig7}), as well as in titles like \textit{Monument Valley}, \textit{Anamorphosis}, \textit{Perfect Angle} and \textit{I Spy Universe}. These models, in particular, highlight the importance of superposition states in explaining aspects of human visual perception~\cite{wang2021game, Mak24_APL}. 

Importantly, the relevance of quantum models extends beyond the domain of video games and virtual reality. In this section, we explore how such models may offer a more powerful and general framework for understanding human cognition and perception in both digital and real-world environments.

The landscape of decision-making research has been marked by intriguing inconsistencies in human choices under risk, often revealing alterations in decisions when subjected to repeated trials \cite{hey1994investigating, blavatskyy2010endowment}. Conventional deterministic decision theories often struggle to account for such alterations \cite{blavatskyy2010models}, which is especially apparent in data obtained from video games \cite{bailey2013would}. The ideas and concepts presented in the preceding sections propose a novel quantum-mechanical model, weaving connections between psychological phenomena and fundamental principles from the realm of physics. Such a model is capable of explaining observed behavioural regularities better than existing decision theories. In particular, video games provide an intriguing platform for exploring decision-making, where human choices under risk reveal surprising inconsistencies and cognitive paradoxes.

\begin{figure}[t]
 \includegraphics[width=1.0\textwidth]{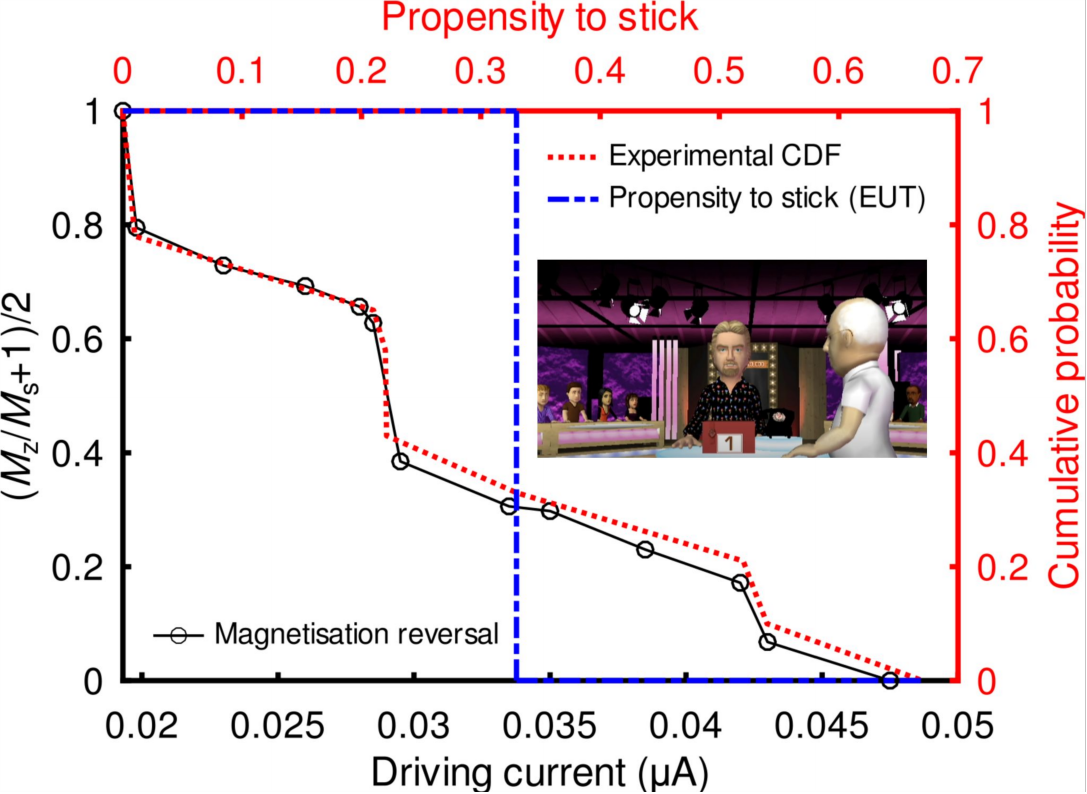}
 \caption{Top $x$ and right $y$ axes:~Cumulative probability distribution function for sticking decisions in the video game. The dashed straight lines denote the EUT propensity to stick. The dotted curve corresponds to the experimental data. Bottom $x$ and left $y$ axes:~The circular markers denote the simulated magnetisation reversal, a purely quantum-mechanical process used in this study to model human decision-making. The thin solid line is the guide to the eye only. The inset shows the interface of the video game used in this study.\label{Fig8}}
\end{figure}

Recent times have witnessed a surge in the popularity of video games, offering a promising avenue for the investigation of decision-making processes within a controlled yet lifelike setting. Among those, the video game `Deal or No Deal' (released in 2006 in the US and in 2008 in the UK by the Gravity-i after the global success of the Endemol~TV show with the same name) stands out as an exemplary terrain for experimental research on decision-making and risk-taking behaviours. The UK version of the video game repeats the `Deal or No Deal UK' show broadcast on Channel~4 (see the inset in Figure~\ref{Fig8}). 

Each game features 22~contestants represented by avatars holding boxes with monetary prizes~\cite{maksymov2025}. The player can select any avatar and any box number and then play the game opening the remaining boxes one by one. Prizes in the game range from \pounds0.01 to \pounds250,000 (hypothetical money). These amounts are randomly distributed across 22~boxes. Once a box is opened and a prize is revealed, it is eliminated from the list of possible prizes. The goal of the participant is to play the game until the end and win the maximum prize of \pounds250,000. Yet, players rarely play the game until the last box and get monetary offers of sure amounts of money from the `banker' after opening specific sequences of boxes: first 5, then 3 repeatedly. Since game players are constantly making choices between a risky lottery and an amount of money for sure (bank offer), the game represents an excellent platform to study human behaviour under risk \cite{pogrebna2008naive,blavatskyy2010models}.

The distinction between sticking and swapping decisions can be evaluated through the methodology used in the field of decision theory, particularly, via the application of Expected Utility Theory (EUT see, e.g., Ref.~\cite{Von44}) and Cumulative Prospect Theory (CPT; see, e.g., Ref.~\cite{Tve92}). Under the deterministic version of EUT, if the expected value of swapping is equal to the expected value of sticking, the model predicts indifference between the two choices.

The term `deterministic decision theories' is standard in decision theory literature and refers to the theories that predict stable choices under identical conditions. The original formulations of EUT and CPT, as well as of the other foundational models, are deterministic in nature, meaning that a decision-maker who prefers option $A$ over $B$ will always make the same choice when faced with the same decision. However, empirical findings in experimental economics and psychology reveal that individuals frequently exhibit choice variability when confronted with repeated decisions, even in identical settings. This discrepancy has led to the development of stochastic models of choice, which introduce probabilistic elements to account for observed behaviour (see, e.g., Refs.~\cite{fechner1948elements, luce1959individual, blavatskyy2010models}).

For instance, while Fechner's model incorporates random errors \cite{fechner1948elements}, Luce's model introduces probabilistic decision rules \cite{luce1959individual} but rank-dependent stochastic models account for response variability. The distinction between deterministic and stochastic choice models is critical since it highlights the need for additional frameworks that can accommodate empirical inconsistencies in human decision-making \cite{hey1994investigating, loomes2017preference}. However, the introduction of stochastic elements, such as Fechner's and Luce's error models, accounts for variability in participant responses across repeated trials. In contrast, CPT posits that loss aversion plays a dominant role in decision-making, leading individuals to overvalue their current choice and disproportionately fear losses associated with swapping. As a result, CPT predicts a stronger preference for sticking to the original box, even when swapping has an objectively equal or higher expected value. Empirical findings from one of our experiments~\cite{Mak24_information1} indicate that while 63\% of participants chose to stick, the observed choice distribution is continuous rather than binary. This suggests that pure decision-theoretic deterministic models do not fully capture decision behaviour, requiring the application of more sophisticated modelling via embedding these theories into models of stochastic choice. This highlights the necessity of integrating stochastic choice mechanisms into theoretical decision-making frameworks \cite{loomes2017preference}.

Historically, studies on decision-making under risk have illuminated the propensity for individuals to modify their choices when confronted with identical binary decision problems repeated over short spans \cite{loomes2017preference}. Traditional deterministic decision theories, rooted in the assumption of consistent choices upon repetition (such as already described deterministic versions of EUT and even Rank-Dependent Utility, of which CPT is a special case), often fall short of explaining these observed deviations \cite{blavatskyy2010models}. Concurrently, a quantum-mechanical perspective has arisen, introducing new possibilities for understanding psychological phenomena through a blend of physics and human behaviour \cite{Bus12, Pot22}. While previous findings highlighted the significance of decision theory and stochastic choice model selection in influencing the estimation of decision theories \cite{blavatskyy2010models}, we demonstrate that synergies between physics and psychology produce more powerful opportunities to explain observed behaviour.

Figure~\ref{Fig8} presents an example of the application of a quantum model---albeit implemented in a distinct mathematical and physical form that involves such well-established quantum-physical concepts as magnetisation~\cite{Mak24_information1}---to data derived from the video game `Deal or No Deal'. The outputs of the quantum model, which are represented by the values of magnetisation and the corresponding current driving the magnetisation dynamics and also associated with the `propensity to stick' in the model, provide an adequate explanatory framework that, to the best of our knowledge, no previous theory has been able to offer.

\section{Quantum-cognitive artificial neural networks}
Modern AI systems---such as large language models (LLMs) that processes and generates human-like text based on input prompts using deep learning techniques to understand context, answer questions, assist with tasks and engage in conversation \cite{Rai24}---are not intelligent in the traditional neurobiology sense. Indeed, rather than possessing true understanding, they are trained on vast datasets and rely on immense computational power to generalise and reassemble information in novel ways based on user prompts.

Emerging systems, often referred to as Agentic AI \cite{Ach25}, represent a step further. These are AI systems capable of acting autonomously with a sense of agency---that is, they can perceive their environment, make decisions, take actions and adapt to achieve specific goals without continuous human oversight. Agentic AI is designed to be proactive, goal-oriented and capable of some reasoning, planning and learning in dynamic and unpredictable contexts.

However, being designed this way does not imply that Agentic AI systems have achieved human-like intelligence. These systems are developed by computer science experts and remain fundamentally software-based, rather than being additionally grounded in knowledge from neuroscience, psychology or decision-making science. 

Quantum models represent one promising approach aimed at bridging this gap, linking advances in computational and data science with insights from neuroscience, psychology and human decision-making. On the other hand, unfortunately, there has been limited communication between the various experts who, ideally, should collaborate on research into quantum cognition and its applications in AI. As a result, as of now, the concept of quantum cognition remains largely confined to specialist circles in the fields of neuroscience, psychology and decision-making science.

Nevertheless, efforts have been made to promote these ideas across disciplinary boundaries~\cite{Bus17, Mak24_APL}. While cognitive AI systems still need to demonstrate clear advantage, preliminary results are promising. Their potential benefit lies in the ability of quantum systems to process information in fundamentally new ways, similar to how quantum computing can outperform classical digital computers:~that is, they are not only faster, but can also tackle classes of problems that digital computers are fundamentally incapable of solving~\cite{Nie02, Mak24_illusions}. Similarly, quantum AI models have been shown to require smaller amounts of training data and can be trained more quickly than their traditionally built counterparts~\cite{Maks25, Maks25_1}.
\begin{figure}[t]
\centering
\includegraphics[width=0.75\columnwidth]{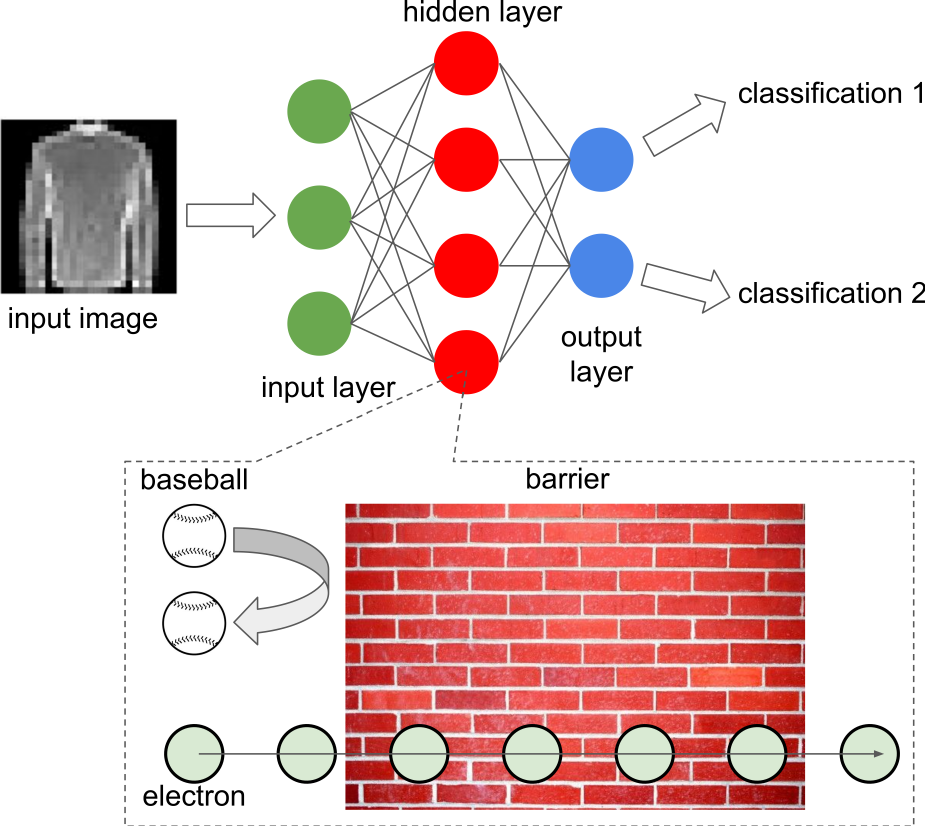}
\caption{Schematic representation of the QT-NN architecture. The inset illustrates the effect of quantum tunnelling that is employed as an activation function of the network.}
\label{Fig9}
\end{figure}

\subsection{Quantum-tunnelling neural networks}
Recently, a novel quantum neural network (QNN) architecture was introduced~\cite{Mak24_APL}, employing the physical phenomenon of quantum tunnelling (QT). As demonstrated in Figure~\ref{Fig2}b and the text discussing its contents, QT describes the transmission of particles through a potential barrier higher than their classical energy~\cite{Gri04}. In classical mechanics, for instance, a baseball with energy $E < V_0$, where $V_0$ denotes the height of the barrier, cannot surmount the obstacle (see Figure~\ref{Fig9}, where the barrier is depicted as a brick wall). In contrast, an electron, being a quantum particle with wave-like properties, exhibits a non-zero probability of penetrating the barrier and continuing its trajectory on the other side (see Figure~\ref{Fig2}b). Even for $E > V_0$, the electron may still be reflected with non-zero probability, due to quantum mechanical effects.

In the QT-based neural network (QT-NN), this phenomenon is mapped onto an electronic circuit, where connection weights are interpreted as the energies of electrons traversing different parts of the network. Crucially, these weights are updated according to quantum mechanical principles. Neurons, in this model, do not require deterministic accumulation of weights for activation. Instead, they operate collectively and probabilistically, introducing additional degrees of freedom and enhancing the model's capacity for processing complex data~\cite{Mak24_APL}.

Although the QT-NN does not use qubits, the incorporation of the tunnelling effect introduces quantum-inspired degrees of freedom during training~\cite{Maks25}. These offer benefits analogous to quantum parallelism and entanglement~\cite{Sho94, Nar00, Nie02}. Moreover, the QT-NN captures uncertainty through the probabilistic nature of its activation functions~\cite{Mak24_APL, Maks25} and through stochastic processes such as white noise injection and random weight initialisation~\cite{Mak24_illusions, Maks25_1}.

What distinguishes the QT-NN from other approaches is its integration of principles from quantum cognition theory~\cite{Atm04, Khr06, Bus12, Pot22}. As demonstrated in the previous sections, this theory merges quantum mechanics with cognitive psychology to model human decision-making and perception. Unlike classical cognitive models that rely on deterministic logic~\cite{Gal_book}, quantum cognition theory states that human behaviour is inherently probabilistic~\cite{Pot22}. It uses concepts such as quantum superposition to explain how individuals may simultaneously hold multiple, even contradictory, beliefs or percepts until a cognitive `measurement' or decision is made~\cite{Bus12} (see Figure~\ref{Fig1}). This framework has proven useful in explaining cognitive phenomena such as optical illusions, biases and decision-making under uncertainty~\cite{Bus12, Mak24_illusions, Mak24_APL, Mak24_information, Mak24_information1}, and presents promising applications in AI systems designed to operate in complex, uncertain environments.

The QT-NN is mathematically grounded in the solution of the Schr{\"o}dinger equation~\cite{Mak24_APL}, a cornerstone of quantum mechanics~\cite{Gri04}. This equation also plays a pivotal role in quantum cognition theory, as its context-dependent solutions mirror the dynamic, probabilistic nature of human thought and perception~\cite{Bus12, Mak24_information}. Furthermore, the QT effect has been organically incorporated into the foundational structure of quantum cognition theory~\cite{Ben18, Mak24_illusions, maksymov2025}, enabling neural network models that more faithfully replicate human cognitive complexity~\cite{Maks25, Maks25_1}.

Recent work has also established strong links between quantum cognition theory and models of consciousness and brain function grounded in quantum information theory~\cite{Geo18, Geo_book, Geo24, Chi19}. As such, the QT-NN architecture functions as a transdisciplinary framework that draws from quantum physics, information theory, psychology, neuroscience and decision science. It enables novel and more realistic modelling of human cognition and decision-making under uncertainty---capabilities that are not achievable with classical models.
\begin{figure}[t]
\centering
\includegraphics[width=0.99\columnwidth]{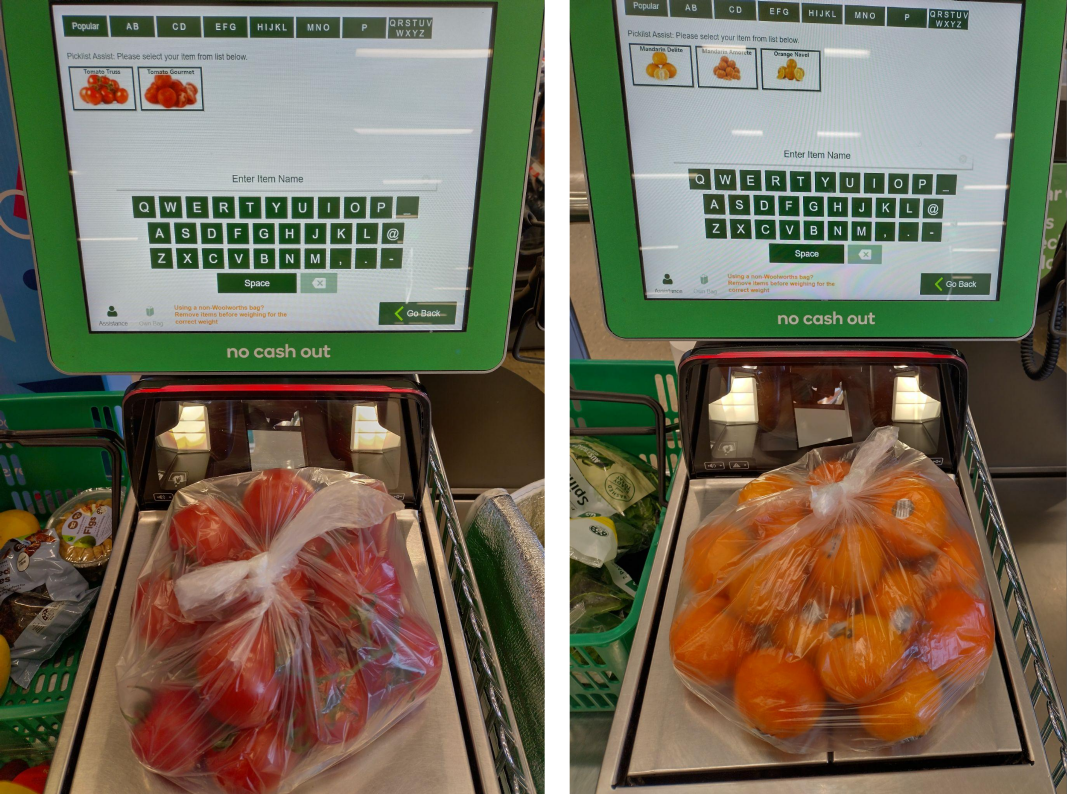}
\caption{Uncertainty in detecting fresh produce items at a supermarket self-checkout equipped with a machine vision system. Left:~The system analysed a transparent plastic bag containing truss tomatoes and identified two possible categories:~truss tomato and gourmet tomato, leaving the final selection to the customer. Right:~In another test with a bag of Amorette mandarins, the system suggested three potential options:~Delite mandarin, Amorette mandarin or Navel orange. Similar results were observed with other visually ambiguous items.}
\label{Fig10}
\end{figure}

\subsection{From optical illusions to real-life objects}
Traditional AI systems are inherently incapable of perceiving optical illusions in the same way as humans, as their underlying algorithms do not account for the psychological and neurological mechanisms of human vision~\cite{Lon04, Kor05, Car14, Kha21}. A growing body of research suggests that this limitation stems from fundamental differences between artificial and biological neurons~\cite{Bak18, Gom20, Kim21, Glo11, Bhu18, Jon21}. While recent studies have increasingly acknowledged this gap~\cite{Wat18, Mel18, Kub21_1, Sha24}, the role of visual and biological complexity in natural environments, and its impact on AI's capacity to replicate human-like perception, remain underexplored~\cite{Pax18, Agr20, Ni20}.

It is worth noting that the study of optical illusions within the context of AI has practical implications that extend far beyond traditional domains such as psychology~\cite{Lon04, Bus12}, the effects of weightlessness~\cite{Yam06, Cle13, Cle17} and video game design~\cite{wang2021game}. As shown above, these investigations offer deeper insights into how AI systems perceive, process and interpret ambiguous information, highlighting their limitations and potential in real-world decision-making scenarios, including those involving different kinds of uncertainty.

Uncertainty is typically defined as the absence of complete knowledge about the current state of a system or the inability to predict future outcomes with precision~\cite{Bus12, Sni19, Gu20}. This concept is especially relevant when dealing with nature, a domain inherently characterised by complexity, nonlinearity and chaotic behaviour~\cite{Str15}. These attributes make accurate forecasting and decision-making difficult and introduce significant epistemic limitations. Figure~\ref{Fig10} illustrates such a scenario using a real-world example.

Furthermore, uncertainty also impedes our understanding of human cognition and behavioural patterns~\cite{Bus12, Sni19}, which in turn can induce anxiety~\cite{Gu20} and erode confidence in high-stakes situations where precise predictions are crucial~\cite{Luc20, Joh23}. Consequently, the rigorous study of uncertainty holds critical importance in machine learning and AI fields of research~\cite{Hul21, Gaw23}.

An important tool in this context is Shannon entropy (SE)~\cite{Hul21, Guh23, Was23}, a central concept in information theory that quantifies the degree of uncertainty associated with a probability distribution~\cite{Kar22}. SE offers a formal mathematical framework for evaluating both the unpredictability and information content of random variables, thereby forging conceptual links between natural and AI~\cite{Sua20}.

In particular, SE serves as a valuable metric for assessing the internal uncertainty of the model and its confidence in particular predictions~\cite{Hul21, Was23}. For example, during neural network training, changes in SE can reflect shifts in weight distributions as the model encounters increasingly complex or ambiguous data~\cite{Hul21, Mak24_APL, Maks25}. Moreover, SE is frequently used to differentiate between confident and uncertain predictions:~lower entropy values generally signify greater model confidence in its classification outcomes~\cite{Was23}. Beyond classification, SE also aids in evaluating the robustness of neural networks with probabilistic layers, especially when exposed to noisy or incomplete data~\cite{Hul21}.

Traditional machine learning models built upon artificial neural networks typically derive connection weights from the limited information encoded in their training datasets~\cite{Fra20}. This data-scarce optimisation often yields models with reduced generalisation capabilities and compromised predictive performance~\cite{Erh10, Jos22, Gaw23}. Furthermore, such models may produce overly confident outputs, despite being inaccurate~\cite{Wan21_1, Wei22}. This overconfidence poses significant risks in safety-critical applications, including autonomous driving~\cite{Mel22}, medical diagnosis~\cite{Wan21_1} and financial decision-making~\cite{Bou23}, where errors can have severe real-world consequences.

To assess the performance of the quantum neural network approach in conditions of uncertainty, we compare the outputs of the QT-NN with those of a classical neural network model trained on the standard Fashion MNIST dataset \cite{Xia17}, which contains 10~classes of fashion items, including shirts, shoes and bags (also see Ref.~\cite{Maks25_1}, where the MNIST dataset of handwritten digits was employed). To ensure a rigorous scientific comparison, both models were constructed with topologically identical architectures, including the same number of neurons, connections and initial random weight distributions, and underwent identical training and testing procedures. 

For instance, when both models were presented with 50 test images from the `Trouser' category of the Fashion MNIST dataset, the equally trained QT-NN and classical model correctly classified them with 93.2\% and 90\% accuracy, respectively. Both models also misclassified some trousers as `Dress', with the QT-NN and classical model assigning 6.8\% and 10\% probabilities to that label, respectively. In another case, the classical model perfectly classified all `Ankle Boot' test images (100\%), while the QT-NN achieved 92.4\% accuracy and made additional, lower-confidence predictions as `Coat', Pullover', Shirt' and `Trouser'.

\begin{figure}[t]
\centering
\includegraphics[width=1.2\columnwidth]{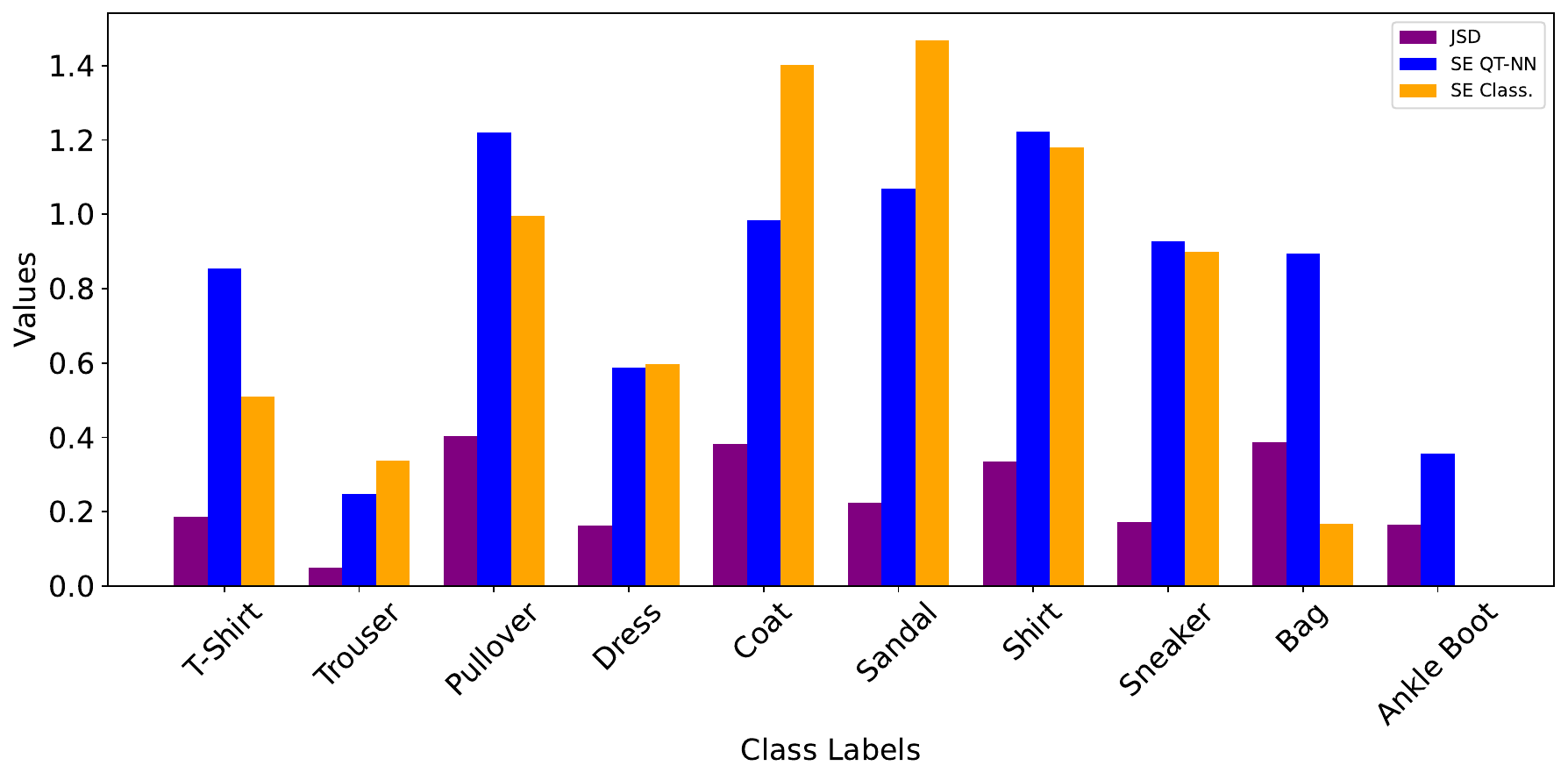}
\caption{JSD and SE figures-of-merit for the QT-NN and classical model across item categories. The classical model’s SE is zero (to machine precision) for the `Ankle Boot' category.}
\label{Fig11}
\end{figure}

To further quantify model performance, we calculated Jensen–Shannon Divergence (JSD) and Shannon Entropy (SE) for each item category (Figure~\ref{Fig11}). A lower JSD indicates higher agreement between the distributions produced by each model. For example, the `Trouser' category yielded a low JSD of 0.049, reflecting near-identical predictions. In contrast, the `Pullover' category exhibited a JSD of 0.403, revealing substantial differences in model interpretations.

Entropy analysis provides complementary insights into prediction uncertainty. Higher SE values indicate more ambiguous or distributed predictions. For example, the QT-NN showed greater uncertainty when classifying `T-Shirt' images (SE = 0.8542) compared to the classical model (SE = 0.5093). Conversely, the classical model exhibited higher uncertainty for categories like `Coat' and `Sandal'. Furthermore, the classical model demonstrated low uncertainty when classifying `Bag' (SE = 0.1672), whereas the QT-NN had a moderate uncertainty level (SE = 0.8943).

Overall, both models demonstrate an ability to accurately classify fashion objects, although they vary in accuracy and uncertainty across categories. In Ref.~\cite{Maks25}, plausible arguments have been presented in support of the claim that the predictions of the QT-NN often align more closely with human perception and cognitive reasoning, especially when considering visual similarities between categories (e.g., `Pullover', `Coat' and `Shirt, or `Dress' and `Trouser'). Furthermore, we can see that footwear categories are consistently and distinctly classified by both models, while `Bag' remains more ambiguous due to its shape similarity to `Coat'. Given its faster training time (see Ref.~\cite{Maks25} for more detail) and interpretability rooted in cognitive reasoning, it can be argued that the QT-NN exhibits superior overall performance. 

\section{Quantum-cognitive AI for drone warfare}
\begin{figure}[t]
\centering
\includegraphics[width=1.2\columnwidth]{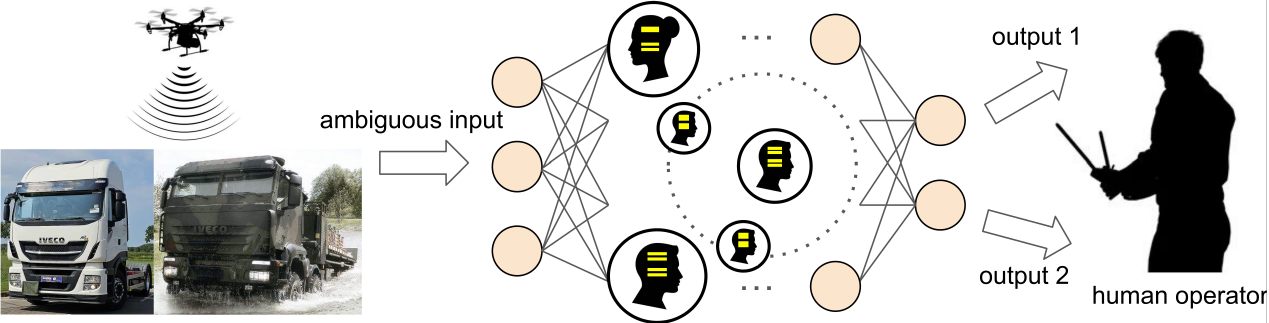}
  \caption{Quantum cognition theory enhances machine learning by incorporating models of human-like bistable perception of optical illusions and cognitive biases into the neural network. The principle of energy quantisation aligns with human mental states (depicted by lines on the head silhouettes, cf.~Figure~\ref{Fig5}), where transitions between energy levels enable nuanced military-civilian vehicle differentiation.}
  \label{Fig12}
\end{figure}

We also demonstrated that QT-based models have the potential to benefit military AI applications by enhancing real-time, adaptive decision-making~\cite{Joh23, Maks25_2, Hum25, Hum25_1}. Through purpose-built test datasets, we show that QT-based Bayesian and recurrent neural networks \cite{Maks25_1}, which combine classical probabilistic models and memory-based approaches with novel QT techniques, improve accuracy in distinguishing military from civilian objects (Figure~\ref{Fig12}) and clarifying verbal commands, thus enabling multimodal quantum-inspired AI and holding the potential to minimise civilian casualties by improving the precision of decision-making in complex, high-pressure environments.

The classification of military and civilian images is a crucial task in training neural network models for military applications \cite{Moy20}. Recent military conflicts have highlighted the growing use of drones in warfare, increasingly targeting vehicles, including civilian ones, driving shifts in combat strategies and operational models \cite{Sur24, Spi25}. The challenge is further compounded by the potential repurposing of civilian vehicles for military operations.

Many military vehicles, particularly European-made trucks \cite{Hei20}, share the chassis and general cabin aesthetics with their civilian counterparts (e.g., Iveco, Tatra and Kamaz multirole vehicles). We found that civilian trucks used in the ongoing military conflicts also appear in the truck subset of the well-known CIFAR-10 dataset \cite{Kri09}. To extend this dataset, we added 1,860 open-source images of custom-identified military vehicles. These images were preprocessed and formatted to be compatible with the CIFAR dataset structure for integration into the training pipeline.
\begin{table}
  \caption{Positive and Negative Words: Military Context}
  \label{tab:military_words}
  \centering
  \begin{tabular}{clcl}
    \toprule
    \multicolumn{2}{c}{\textbf{Positive Words}} & \multicolumn{2}{c}{\textbf{Negative Words}} \\ 
    \midrule
    Achieve       & Advance       & Abort       & Ambiguous \\ 
    Authorize     & Clear         & Breakdown   & Cancel \\ 
    Command       & Confirm       & Compromised & Conflicted \\ 
    Decisive      & Definitive    & Degrade     & Defeat \\ 
    Deploy        & Designated    & Denied      & Disrupt \\ 
    Effective     & Engage        & Doubtful    & Failure \\ 
    Established   & Mission-ready & Ineffective & Misfire \\ 
    Objective-secured & On-target & Obstructed  & Off-course \\ 
    Success       & Validated     & Unconfirmed & Void \\ 
    \bottomrule
  \end{tabular}
\end{table}

To showcase the potential of the proposed QT-based models for use in multimodal solutions, we also created a custom database of terms commonly used in operational planning, rules of engagement and mission status reports and tested in using the QT-based recurrent neural network model. These words reflect either successful or unsuccessful outcomes when defining and confirming objectives in a military context (Table~\ref{tab:military_words}). They were arranged into meaningful phrases that mimic the communication between the operators of military drones.

As a result, both the prior work \cite{Maks25, Maks25_1} and this study~\cite{Maks25_2} provide strong evidence supporting the ability of the QT model to mimic human-like perception and judgement in decision-making. The statistical model developed in Ref.~\cite{Maks25} also demonstrated that the QT model classifies images more flexibly than traditional approaches and aligns more closely with human cognitive processes. Comparisons of outputs, weight distributions and uncertainty measures further confirmed that the QT model trains more efficiently and delivers comparable or improved accuracy relative to similarly configured conventional models. Nevertheless, further human trials are required for definitive validation.

Subsequent investigations will also involve further calibration of the model output using human-produced experimental data. In the QT model, the hyperparameters include the width and height of the potential barrier, which control the structure of the discrete quantum energy levels (see Figure~\ref{Fig5} and Refs.~\cite{Maks25, Maks25_1}). Since these energy levels have been linked to human mental states \cite{Mak24_information}, theoretical progress can be made by adjusting the quantum behaviour of the model to more closely align with human cognitive and emotional responses.

Future work within the emphasised directions will also advance a three-level ethical approach to military AI, comprising data, performance and values. This situates the proposed model within the inherent logic of responsible military AI \cite{Anna1}, thereby encompassing considerations such as transparency and explainability, privacy and data protection, robustness and safety, and inclusion and anti-discrimination---all of which collectively impact notions of trust, accountability and fairness \cite{Anna2, Sur24_1}.

By integrating both human and machine capabilities into the model-building process, the approach proposed in this work can offer improved explainability. It has the potential to provide insights into the sophistication of human decision-making, including judgements considered to be `common sense', which the model seeks to approximate~\cite{Hum25}. This, in turn, reinforces human-centricity as a core value in AI development \cite{Anna5}. Moreover, it supports the previously articulated view that differences in machine learning methodologies can influence a model’s ability to align with the principles of responsible AI \cite{Anna6}. Generally speaking, this study renders concerns about fairness and inclusion less prominent, as the model explicitly incorporates elements of human cognition involved in visual recognition, such as pattern recognition, categorisation and abstract reasoning. Importantly, the findings are expected to deepen our understanding of cognition within the context of human–machine teaming, particularly in contemporary military operations.

\section{The quantum turn: science, cognition and possibility}
To conclude this chapter, we will try to address a frequently posed question:~`Why quantum physics?' Providing a definitive response to this inquiry presents a greater challenge than might initially be assumed. Indeed, the preceding discussion has been fundamentally concerned with the application of quantum physics to elucidate socio-political phenomena and augment the cognitive capacities of AI systems---with demonstrably promising results. A plausible answer may be manifold. 

Firstly, quantum physicists, as well as individuals working in the adjacent fields such as mathematics and engineering, are uniquely positioned to drive progress in quantum cognition theory because their expertise bridges the abstract and the empirically testable~\cite{Sur21, Mel23, Sur24_1}. At its core, quantum cognition borrows mathematical tools from quantum mechanics---such as superposition, entanglement and interference~\cite{Khr06, Bus12, Pot22}---to model complex cognitive phenomena that defy classical explanation. These include ambiguous decision-making, context-dependent judgements and many aspects of machine learning. However, applying these principles correctly requires more than just metaphorical parallels; it demands a rigorous grasp of quantum formalism. Physicists, trained to understand such concepts as Hilbert spaces, non-commutative operations and probabilistic wavefunctions, can ensure the models remain mathematically sound rather than merely suggestive. Without this foundation, quantum cognition risks becoming a loose analogy rather than a predictive framework.

Moreover, quantum physicists and expert with relevant training bring a discipline of first-principles thinking that is critical for innovation in this interdisciplinary space. They are adept at translating between theory and experiment, a skill vital for refining quantum cognitive models into testable hypotheses. While psychologists and AI researchers provide essential insights into human and machine behaviour, physicists anchor the work in quantifiable mechanics, preventing conceptual drift. Their ability to spot when quantum effects truly matter (and when they do not) helps the field avoid overextension. Ultimately, advancing quantum cognition is not just about borrowing quantum ideas; it is about rigorously adapting them. And that is a task best led by those who speak the language fluently. 

Secondly, while a digital computer executing the algorithms presented in this text remains fundamentally classical---despite its quantum-inspired mathematical framework (Schr{\"o}dinger equation solutions) and quantum-generated random inputs~\cite{Mak24_illusions}---this classical implementation is immaterial within sociophysics~\cite{Gal_book} and quantum cognition theory~\cite{Bus12, Pot22}. These frameworks prioritise model validity over physical substrate, provided the underlying idealisations and data assumptions hold empirically~\cite{Mak24_information}. This principle is additionally exemplified by classical electrodynamical models that successfully replicate human preference imprecision using quantum-derived phenomena like spin-transfer torque \cite{Mak24_information1}.

The classical-quantum distinction gains significance, however, in theories linking quantum information processes to consciousness and brain function \cite{Chi19, Geo_book}. While quantum cognition and quantum brain theories remain formally distinct \cite{Bus12, Pot22}, their eventual convergence toward a unified framework appears inevitable \cite{Sch05_1, Bus17, Khr20}. Though such synthesis lies beyond this work's scope, we emphasise that a hardware-based implementation of the algorithms and models presented above would constitute a true quantum neuromorphic system, an analog platform distinct from conventional quantum computing architectures \cite{Mar20_2}. Critically, this physical realisation would meet the `genuine quantum hardware' criterion \cite{Chi19} deemed essential in quantum brain theories for modelling consciousness \cite{Geo_book}.

Finally, research on the perception of optical illusions demonstrates that quantum physics can provide insights into complex socio-cognitive phenomena, including gender fluidity \cite{Tac20, Mak24_gender}. This intersection further enriches the transdisciplinary framework emphasised in this study.

\section*{Acknowledgements}
The author would like to thank his collaborators Ganna Pogrebna, A.~H.~Abbas, Milan Maksimovic, Anna Bohdanets, Immaculate Motsi-Omoijiade and Guido Governatori for fruitful discussions that helped shape different parts this work.

\section*{Data availability}
The source codes that implement the neural network models discussed in this document are available in the GitHub repository, \url{https://github.com/IvanMaksymov/Quantum-Tunnelling-Neural-Networks-Tutorial} and \url{https://github.com/IvanMaksymov/QT-CIFAR-Military}.

\bibliographystyle{sn-aps}
\bibliography{sample}


\end{document}